\documentclass[twocolumn,superscriptaddress,pra,amsmath,amssymb]{revtex4}

\usepackage{graphicx}

\newcommand{\diff}{\mathrm{d}}

\newcommand{\erf}{\text{\rm erf}}

\newcommand{\vx}{\boldsymbol{x}}

\newcommand{\oM}{\mathsf m }

\newcommand{\on}{ \mathsf n }
\newcommand{\ovX}{ \mathsf {X} }

\newcommand{\V}{ {\cal V} }
\newcommand{\T}{ {\cal T} }

\newcommand{\rc}{r_{\rm c}}
\newcommand{\Vsin}{\V}
\newcommand{\Vsincsl}{\V_{\rm CSL}}

\begin{document}

\title{Testing spontaneous localization theories with matter-wave interferometry}

\author{Stefan Nimmrichter}
\affiliation{University of Vienna, Vienna Center for Quantum Science and Technology (VCQ), Faculty of Physics, Boltzmanngasse 5, 1090 Vienna, Austria}

\author{Klaus Hornberger}
\affiliation{University of Duisburg-Essen, Faculty of Physics, Lotharstra{\ss}e 1-21, 47048 Duisburg, Germany}
\affiliation{Max-Planck Institute for the Physics of Complex Systems, N\"{o}thnitzer Stra{\ss}e 38, 01187 Dresden, Germany}

\author{Philipp Haslinger}
\author{Markus Arndt}
\affiliation{University of Vienna, Vienna Center for Quantum Science and Technology (VCQ), Faculty of Physics, Boltzmanngasse 5, 1090 Vienna, Austria}

\date{\today}

\begin{abstract}
We propose to test the theory of continuous spontaneous localization (CSL) in an all-optical time-domain Talbot-Lau interferometer for clusters with masses exceeding $10^6 \,$amu. By assessing the relevant environmental decoherence mechanisms, as well as the growing size of the particles relative to the grating fringes, we argue that it will be feasible to test the quantum superposition principle in a mass range excluded by recent estimates of the CSL effect.
\end{abstract}

\maketitle

\section{Introduction}

It is a basic unresolved question of quantum mechanics 
whether the Schr{\"o}dinger equation holds for truly macroscopic systems. Unitarity then implies that even measurement devices or conscious observers can, in principle, be brought into a superposition of macroscopically distinct states.
Most of the offered answers can be put into one of three categories.
The affirmative statement, preferred in quantum cosmology, requires some interpretational exercise to explain why definite  measurement outcomes are perceived in spite of all outcomes being simultaneously realized in a multitude of `Everett worlds'. A quite different attitude, expressed most stringently in the operationalist formulation of quantum mechanics, attributes a fundamental role to the divide between `quantum system' and measurement device such that it is meaningless to pose the question in the first place. The third option is to hypothesize that there is an objective modification of the unitary Schr{\"o}dinger dynamics, which gives rise to a macrorealist description of the physics on macroscopic scales \cite{Leggett2002}.

Whatever one thinks about the need or plausibility of such unconventional theories of the quantum-to-classical transition, they have the clear advantage that they can be tested in principle. This way they bring back to physics what is otherwise an issue of logical consistency and epistemology. Another motivation to consider the possibility that quantum physics is only an approximation to a deeper underlying theory may be drawn from the difficulties encountered when tries to reconcile it with the theory of gravity \cite{Penrose2004road}.

One of the best studied models for the emergence of macrorealism is the theory of Ghirardi, Rimini and Weber (GRW) \cite{Ghirardi1986}, and its refinement, the theory of continuous spontaneous localization (CSL) \cite{CSLboth}.
Its predictions are consistent with all quantum experiments so far, but they strongly deviate from quantum theory when applied to macroscopic objects \cite{Bassi2003}. According to the model, a delocalized quantum state of a material particle may experience a random ``collapse'', which localizes the wave function to a scale of about $100\,$nm. In case of composite objects the rate of these collapse events increases with mass due to an inherent amplification mechanism. The values of the localization parameters are chosen such that they affect only systems considered to be in the macroscopic domain.

In this letter, we propose to test the CSL model by performing matter wave interference with clusters in the mass range between $10^6$ and $10^8$\,amu. 
We assess the various relevant environmental decoherence processes expected to occur in an optimized time-domain Talbot-Lau interferometer with ultra-violet laser gratings, and in particular the enhanced signal loss due to the finite cluster size.
We conclude that it will become technologically feasible to test the quantum superposition principle at mass and time scales where it is predicted to fail according to recent estimates by Adler and Bassi \cite{AdlerBassiCSL,Bassi2010}.

\section{Effects of continuous spontaneous localization}

The observable consequences of the CSL model are accounted for in the framework of second quantization by adding the Lindblad term
\begin{equation}
\frac{8 \pi^{3/2}\rc^3 \lambda_0}{m_0^2} \! \int \!\! \diff \vx \left[ \oM(\vx) \varrho \, \oM(\vx) - \tfrac{1}{2} \left\{ \oM^2 (\vx),\varrho \right\} \right] \label{eqn:CSL}
\end{equation} 
to the von Neumann equation for the many-particle density operator $\varrho$ of a system of massive particles.
Here, $ \oM(\vx) $ is the spatially blurred mass density operator, defined in terms of the number density operators $\on_k(\vx) $ of the constituent boson and fermion species and their respective masses $m_k$,
\begin{equation}
 \oM (\vx) = \int \!\! \diff \vx' \, g(\vx-\vx') \sum_{k} m_k \on_k (\vx'). \label{eqn:M}
\end{equation}
The function $g(\vx)$ is a normalized gaussian whose width $\rc=100$\,nm is one of the parameters of the model. The second parameter is the term $\lambda_0/m_0^2$. One conventionally chooses the reference mass $m_0$ to be given by a nucleon, $m_0=1$\,amu; as discussed in \cite{AdlerBassiCSL,Bassi2010}, reasonable lower bounds
for the associated localization rate $\lambda_0$ are then in the range of
$10^{-8}$\,Hz to $10^{-12}$\,Hz. This is substantially larger than the value of $10^{-16}$\,Hz originally suggested by GRW \cite{Ghirardi1986}, since the CSL model implies a quadratic mass dependence of the effective localization rate. It is a consequence of the second quantization formulation, which guarantees that the exchange symmetry of bosons and fermions remains unaffected by the collapse events.

In a molecule or a cluster, where the inter-particle distances are much smaller than the localization scale $\rc$,
the collapses affect only the quantum state $\rho$ of the center of mass motion. One arrives at an effective master equation $\partial_t\rho=(i\hbar)^{-1}[\mathsf{H},\rho]+ \mathcal{L}\rho$ with
\begin{equation}
 \mathcal{L}\rho = \lambda \left[
8\pi^{3/2}\rc^3
\! \int \!\! \diff \vx' \, g\left( \ovX - \vx' \right) \rho \, g\left( \ovX - \vx' \right) - \rho \right] \label{eqn:CSLcom}.
\end{equation} 
Here $\ovX$ is the center-of-mass position operator, and $\lambda = \lambda_0 (m/m_0)^2$ the effective localization rate depending quadratically on the total mass $m$ \cite{AdlerBassiCSL,Bassi2010}. 

We note that the master equation (\ref{eqn:CSLcom}) is equivalent to a collisional decoherence master equation \cite{Vacchini2007b}. This implies that it induces a basis of localized, soliton-like pointer states for sufficiently large $\lambda$, which move without dispersion on the classical Newtonian trajectories \cite{Busse2010a}.

\section{Testing spontaneous localization with near field interference}

We learn from Eq.~(\ref{eqn:CSLcom}) that continuous spontaneous localization can be tested as soon as a very massive particle is brought into a superposition state of different positions which exceed the distance of $\rc=100$\,nm for a sufficiently long time. 
We propose that a viable experiment can be based on the optical time-domain ionizing matter (OTIMA) interferometer described in \cite{optimapreprint}. 

In that experiment, a pulsed slow cloud of clusters is subjected to three pulsed standing light waves, generated for instance by an ultra violet fluorine laser beam ($\lambda_{\rm L} = 157\,$nm), such that the particles in the antinodes are ionized and removed from the cloud. The number of remaining neutral clusters is then recorded as a function of the delay times between the grating pulses. 
This way the first laser pulse generates spatial coherence in the cluster cloud by modulating its initial density. After a delay time $T$ the second pulse acts as a combined absorption and phase grating, where the nodes of the standing light field play the role of the `grating slits' with a period of $d=\lambda_{\rm L}/2$, while the phases of the matter waves get shifted by the dispersive light-matter interaction.
Talbot-Lau-type near field interference finally produces a periodic cluster density pattern after a second time delay $T$, provided $T$ is close to an integer multiple of the Talbot time $T_{\rm T} = md^2/h$,
i.e., $T = N T_{\rm T} + \delta T$ with $\delta T/T\ll 1$.
The recorded signal is predicted \cite{optimapreprint} to show high-contrast interference fringes
as a function of $\delta T$, which are conveniently characterized by the sinusoidal visibility $\Vsin$ defined as the ratio of amplitude and offset of a fitted sine curve.

\begin{figure}
\includegraphics[width=\columnwidth]{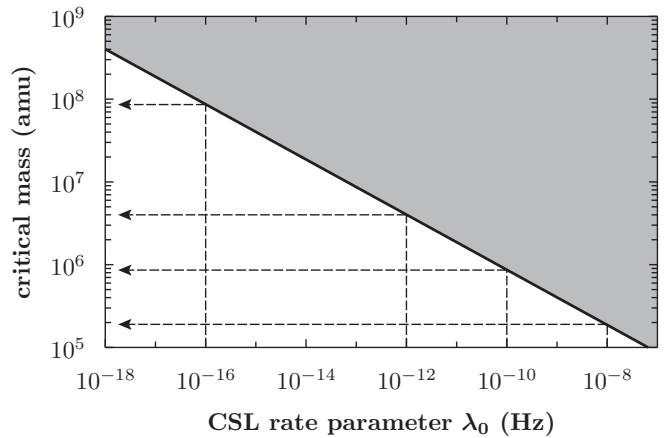}
\caption{\label{fig:csl}
Critical mass $m_{\rm c}$ for testing continuous spontaneous localization as a function of the localization rate $\lambda_0$, assuming the geometry of the proposed experiment with $N=2$. The shaded area indicates the parameter region where interference should be unobservable according to the CSL model. The dashed arrows mark the critical masses associated to reasonable estimates for the lower bound of $\lambda_0$ \cite{AdlerBassiCSL,Bassi2010}.}
\end{figure}

If continuous spontaneous localization exists we predict a reduction of the interference visibility, which can be calculated by incorporating the CSL master equation (\ref{eqn:CSLcom}) into the theoretical description of the interferometer, like in the case of environmental decoherence \cite{Hornberger2004}. One arrives at the closed expression
\begin{equation}
\frac{\Vsincsl}{\Vsin} = \exp\left\{-2 \lambda_0 T_0 N  \frac{m^3}{m_0^3} \! \left[1- \frac{
\sqrt{\pi}\rc}{Nd} \erf\left( \frac{N d}{2 \rc} \right) \right]  \right\} . \label{eqn:RCSL}
\end{equation} 
Here we introduced the Talbot time per atomic mass unit $T_0 = m_0 d^2/h$. The cubic mass dependence in the exponent is the reason why an interferometric test of the CSL model becomes conceivable.

Figure \ref{fig:csl} shows the critical cluster mass $m_{\rm c}$  where continuous spontaneous localization predicts the reduction
$\Vsincsl=\Vsin/2$ in the proposed OTIMA experiment. Observing substantial interference at $m_c$ thus puts a strict upper bound on the localization rate $\lambda_0$ by ruling out values
greater than those given by the solid line.
Here we take $N=2$, i.e.~the second Talbot order. This ensures that the
effective interference path separation $N d=157\,$nm exceeds the localization scale $\rc$, such that the factor in square brackets takes an appreciable value.

Using Fig.~\ref{fig:csl} we can assess whether the literature values for the CSL rate can be tested in the proposed setup. The most recent calculations
suggest $\lambda_0=10^{-10 \pm 2}\,$Hz \cite{AdlerBassiCSL,Bassi2010}, while the original GRW estimate gives $\lambda_0 = 10^{-16}\,$Hz \cite{Ghirardi1986}. 
These values require the cluster mass to reach $m_{\rm c}= 10^{5.9 \pm 0.7}\,$amu and $m_{\rm c}=9\times 10^7\,$amu, respectively, as marked by the dashed arrows in Fig.~\ref{fig:csl}. Remarkably, the lower value already touches the range which should be reached with present-day technology in the OTIMA interferometer \cite{optimapreprint}. However, for a distinctive test of continuous spontaneous localization one must venture beyond that, which requires to cope in particular with two additional types of contrast limiting effects \footnote{%
In A.~Bassi, E.~Ippoliti, and S.~Adler, Phys. Rev. Lett. {\bf 94}, 030401 (2005), it was suggested to test CSL in a superposition experiment with a mirror, which would have to be cooled to its vibrational ground state and to be sufficiently isolated to prevent environmental decoherence. If $50\%$ of the coherence could be maintained during one oscillation period, this would imply $\lambda_0 < 7 \times 10^{-9}\,$Hz, which touches only the upper end of the recent estimates.%
}.

First, the total interference time $2N T_{\rm T}$ grows linearly with the cluster mass. At $10^6\,$amu it already amounts to about $60\,$ms at the second Talbot order, if a 157\,nm laser is used. 
At $10^8\,$amu the gravitational free fall must be compensated and the cluster velocity must be controlled with high precision. This requires motional slowing,  guiding \cite{Schiffer1998}, and trapping techniques \cite{Cohen2005,Chang2010} for large clusters, and possibly a microgravitational environment \cite{Selig2010}. The increased interference time also aggravates the decohering influence of environmental interactions and external forces, as discussed in the following section.

Second, the size of the clusters grows with increasing mass.
Apart from enhancing the interaction with the radiation field and with residual gas particles, this implies that the cluster size becomes comparable to the period $d=\lambda_{\rm L}/2$ of the optical gratings. 
Starting with this last problem, we proceed to analyze under what conditions OTIMA interference will be possible with ultra-massive clusters, so as to provide a testing ground for spontaneous localization.

\section{OTIMA interference in the limit of high particle masses}

In the molecular matter-wave interference experiments carried out so far, it was justified to treat the delocalized objects as polarizable point particles.
Since this approximation breaks down beyond $10^6\,$amu, we formulate the interaction between the ionizing optical gratings and the finite-size spherical clusters using Mie theory \cite{Mieboth}.
As explained in \cite{optimapreprint}, both coherent diffraction and photoabsorption are described by $n(x)$, the average number of photons absorbed by each cluster during a laser pulse,
\begin{equation}
 n(x) = n_{0} + n_{1} \cos \left( \frac{2 \pi x}{d} \right). \label{eqn:nx}
\end{equation} 
Here, $x$ is the transverse center-of-mass position of the cluster, and $n_{0}$ is the  position-averaged mean number of absorbed photons. Only the modulation $n_{1}$ gives rise to interference; the corresponding visibility \cite{optimapreprint}
\begin{equation}
 \Vsin = 2 \frac{I_1^2 \left( n_{1} \right) I_2 \left( n_{1} \right) }{ I_0^3 \left( n_{1} \right) } \label{eqn:Vsin}
\end{equation} 
is independent of the Talbot order $N$, and involves modified Bessel functions of the first kind. The position average $n_{0}$, on the other hand, determines the total transmissivity $\T$ of the three gratings, i.e.~the fraction of remaining neutral clusters after three grating pulses,
\begin{equation}
 \T = \exp \left( -3 n_{0} \right) I_0^3 \left( n_{1} \right). \label{eqn:Trans}
\end{equation} 
For clusters small compared to the laser wave length the point particle approximation yields $n_{1}=n_{0}=2F_{\rm L} \sigma_{\rm abs} / h \nu_{\rm L}$, proportional to the absorption cross section and to the energy flux $F_{\rm L}$ of the running-wave laser input \cite{optimapreprint}.
For large clusters the corresponding expressions  can be obtained by computing the absorbed power of a dielectric sphere with radius $R$ in a standing-wave field. A lengthy calculation yields
\begin{equation}
\label{eq:n01}
n_{\substack{1\\0}}
= \frac{4 F_{\rm L}}{h\nu_{\rm L}} \sum_{\ell = 1}^\infty \frac{(2\ell+1)\pi}{k_{\rm L}^2 \rho} \left( \mp \right)^{\ell-1} \left( \sigma_{\ell}^{(E)} \pm \sigma_{\ell}^{(H)} \right),
\end{equation}
with the electric and magnetic multipole components
\begin{eqnarray}
 \sigma_{\ell}^{(E)} &=& \frac{{\rm Im} \left\{ \varepsilon j_{\ell} \left[ \sqrt{\varepsilon} \rho j_{\ell-1} - \ell j_{\ell} \right]^{*}  \right\} }{ \left| \ell (\varepsilon-1)j_{\ell} h_{\ell} + \sqrt{\varepsilon} \rho \left[ j_{\ell-1} h_{\ell} - \sqrt{\varepsilon} j_{\ell} h_{\ell-1} \right] \right|^2} ,\nonumber\\%
\label{eq:sigmas}
 \sigma_{\ell}^{(H)} &=& \frac{{\rm Im} \left\{ \sqrt{\varepsilon} j_{\ell}^{*} j_{\ell - 1}  \right\} }{ \rho \left| j_{\ell} h_{\ell+1} - \sqrt{\varepsilon} j_{\ell+1} h_{\ell} \right|^2}.
\end{eqnarray}
The latter are determined by the scaled cluster radius $\rho = k_{\rm L} R = 2\pi R/\lambda_{\rm L}$ and by the relative permittivity $\varepsilon$ of the material. We use a short-hand notation for the spherical Bessel function in the dielectric, $j_{\ell} \equiv j_{\ell} \left( \sqrt{\varepsilon} \rho \right)$, and for the vacuum solutions $h_{\ell} \equiv h_{\ell}^{(1)} (\rho)$, given by spherical Hankel functions of the first kind.

As the clusters grow in size, both absorption parameters $n_{0}$ and $n_{1}$ increase. This can be compensated by turning down the grating laser flux $F_{\rm L}$. However, if we want to maintain a fixed visibility (\ref{eqn:Vsin}) for growing clusters, the detection probability per particle $\T$ drops rapidly. This is because the ratio $n_{0}/n_{1}$ grows drastically once the cluster radius becomes comparable to the grating period. That is the main effect of the cluster size in the subwavelength regime. 

\begin{figure}
\includegraphics[width=\columnwidth]{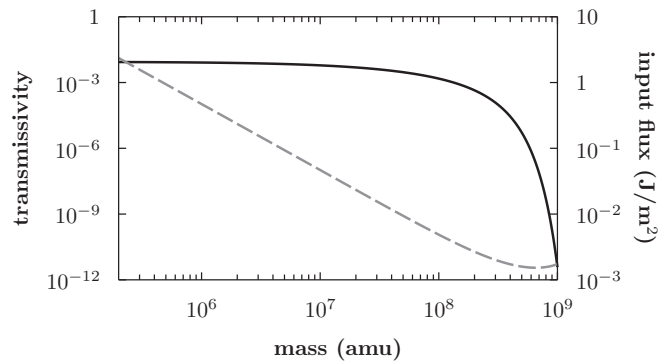}
\caption{\label{fig:size} Transmissivity (\ref{eqn:Trans}) of the  OTIMA interferometer versus the gold cluster mass.
The dashed line (right scale) gives the laser flux, required to fix the visibility at $\Vsin=85\%$. }
\end{figure}

Figure~\ref{fig:size} shows the OTIMA transmissivity for gold clusters as a function of their mass.
The relative permittivity at $\lambda_{\rm L} = 157\,$nm is approximated by the bulk value $\varepsilon = 0.9 + 3.2i$, 
as is the mass density.
We keep the interference fringe visibility (\ref{eqn:Vsin}) at $85\%$ by varying the laser flux {\em (dashed line)}. 
When comparing the medium-sized Au$_{1000}$ with a 1000 times more massive cluster, an almost 1000 times weaker laser pulse will suffice for the latter. The transmissivity, however, drops then from $\T = 1\%$ to $4 \times 10^{-4}$ of the incident cluster flux. This shows that the signal loss becomes prohibitively large beyond $10^8\,$amu, where cluster size and wavelength become comparable even for the densest metals. 

\begin{figure}
\includegraphics[width=\columnwidth]{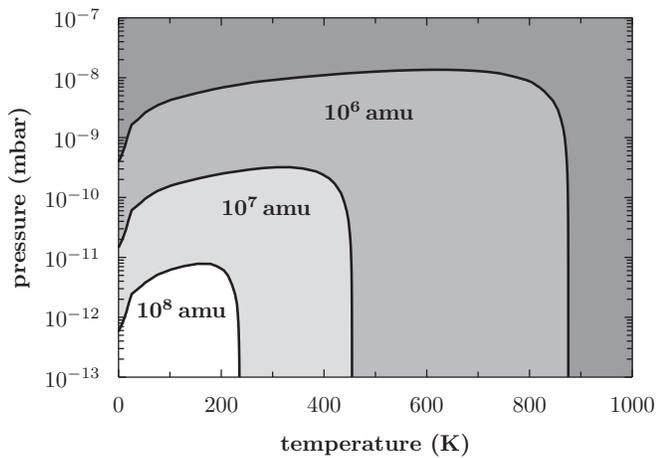}
\caption{\label{fig:decoherence}
The contour lines give the critical residual gas pressures and the critical ambient temperatures for observing interference of gold clusters with masses of $10^6\,$amu, $10^7\,$amu,  and $10^8\,$amu.}
\end{figure}

However, even before this size-induced mass limit is reached, environmental decoherence will become important if the cluster particles are not properly isolated and cooled. We identify three relevant decoherence processes, {\em(i)} the elastic scattering of laser photons during the grating pulses, {\em(ii)} collisions with residual gas particles, and {\em(iii)} the emission, absorption and scattering of thermal blackbody photons. The first process was shown \cite{optimapreprint} to be negligible below $10^9\,$amu. The other two effects pose constraints on the background pressure and the temperature of the setup. Both effects are well understood \cite{Hornberger2004,Hackermuller2003b, Hackermueller2004a}. Their dependence on the cluster mass differs from the continuous spontaneous localization effect (\ref{eqn:RCSL}), and allows one in principle to separate the environmentally induced visibility reduction from CSL by independently varying the temperature and the pressure of the interferometer chamber.

In Figure \ref{fig:decoherence} we plot the critical background pressure $p_{\rm c}$ and temperature $T_{\rm c}$, where environmental decoherence 
reduces the interference visibility by a factor of two. The contour lines correspond to the cluster masses $m/{\rm amu}=10^6$, $10^7$, and $10^8$, with the shaded area indicating where interference is largely suppressed by the environment. The area of high fringe visibility shrinks with growing cluster mass since the optical absorption, the Rayleigh scattering cross section, and the collisional cross section all grow with size. Here we assume the residual gas to consist of N$_2$, and we use the electric properties of bulk gold to assess all decoherence processes. 
The cluster temperature is assumed to be in equilibrium with the environment, implying that in addition to the thermal emission of radiation both the absorption and the elastic scattering of blackbody photons contribute significantly to the thermal decoherence effect.

We infer from Fig.~\ref{fig:decoherence} that decoherence can be fairly easily controlled for $10^6\,$amu clusters at a pressure of $10^{-9}\,$mbar and room temperature.
An experiment with $10^8\,$amu, in contrast, will require to cool the setup and the clusters to below $200\,$K, in a chamber evacuated below $10^{-12}\,$mbar. 
This is challenging but feasible, as demonstrated by cryogenic trap experiments at $10^{-17}$\,mbar and at 4.2\,K \cite{Gabrielse1990}.

\section{Conclusions}

For a long time the implications of the theory of continuous spontaneous localization were thought to be practically unobservable. Our present assessment shows that this is not the case. Indeed, experiments aimed at demonstrating matter-wave interference with massive clusters in the range between $10^6$ and $10^8$\,amu will provide an ideal testing ground for this unconventional theory of the quantum-to-classical transition, one of the leading contenders in resolving the fundamental question of macroscopic realism.

\acknowledgments
We thank the Austrian Science Fund (FWF) for support within
the projects Wittgenstein Z149-N16 and  DK-W1210 CoQuS.
We also acknowledge financial support within the \emph{MIME} project of the ESF EuroQUASAR program, through FWF and Deutsche Forschungsgemeinschaft (DFG).

\end{document}